\documentstyle[12pt,a4,doublespace,epsfig]{article}
\begin{document}
%\hyphenation{http://DUR-PDG.DUR.AC.UK/HEP-DATA}
\title{Determination of the forward Compton scattering amplitudes for C and Pb}
\author{A. Deppman\thanks{partially supported by Conselho Nacional de
Desenvolvimento Cient\'\i fico e Tecnol\'ogico - CNPq, Brazil.},
N. Bianchi, E. De Sanctis\thanks{DESANCTIS@LNF.INFN.IT},\\ 
M. Mirazita, V. Muccifora, P. Rossi\\
Laboratori Nazionali di Frascati,\\
Istituto Nazionale di Fisica Nucleare,\\
P.O.Box 13, 00044 Frascati (Roma), Italy.}
%\date{}
\maketitle

\begin{abstract}
The forward Compton scattering amplitudes for carbon and 
lead have been calculated from  total photoabsorption cross 
section data by using
dispersion relations. The results show a large difference between the
scattering amplitudes 
for nuclei and the free nucleon, above the
$\Delta$ region.
Difference between carbon and lead is also evident.
The forward Compton scattering cross sections have been calculated 
and compared with the available data. The Weise sum rule is discussed
together with the predictions of a recent theoretical 
model.
\end{abstract}


\section{Introduction}

Intermediate energy photonuclear reactions have been intensively studied in 
the last few years to investigate the nucleon's properties in nuclear 
matter. In this context, the spin averaged Forward Compton scattering
Amplitude (FCA) is an interesting quantity, since it is directly related to 
the 
scattered wave function and to the scattering potential. In addition, it allows
to connect photon-nucleon and photon-nucleus cross sections through a 
generalized photoabsorption sum rule which reconciles the enhancement of the
low-energy 
integrated photonuclear cross section over the classical dipole sum
with the shadowing behaviour observed at high-energy.

In 1988, Ahrens et al.~\cite{AHR} performed a dispersion relation analysis
of the photoabsorption cross section. Due to the lack of nuclear data 
at intermediate energies (from about~0.5 to about~2~GeV), their analysis
was accurate to energies up to the $\Delta$ resonance region. They found that
the real part of the 
nuclear FCA has a universal zero at the photon energy
$\omega =$~327~$\pm$~5~MeV 
throughout the periodic table, and close to the zero for the free proton, 
found at $\omega=$~318~MeV. This fact was interpreted as a clear evidence 
that the $\Delta$-resonance survives as a quasi-particle in nuclei with 
mass shifted by only a small amount from the free $\Delta$-resonance mass.
On the contrary recent measurement of the total photoabsorption cross 
sections for several nuclei in the energy region 300-1200 MeV has shown that
the $\Delta$-resonance mass and width increase almost linearly with the 
nuclear density~\cite{BIAPB}.

The topic of the hadronic mass modifications in the nuclear  
medium has received intense interest of the theoretical community. For a review 
see reference~\cite{HAT}.
In particular, it is worth noticing that the nuclear medium dependence
of the baryon mass implies a change in the properties 
of the vacuum which translates into a scaling (density and/or temperature
dependence) of the free model paramenters. In this context recently Mukhopadhyay
 and Vento~\cite{MUK}
have analysed  the baryon and meson mass shifts 
in nuclei in the framework of two different QCD-inspired models, 
offering for the first time different insights and 
connections in a subject well known in the 
traditional nuclear physics domain. In particular their investigation has shown
that the baryon mass shift suggests a partial quark deconfinement in the 
nuclear medium.

Recently, Boffi et al.~\cite{BOFFI} used the
Weise sum rule~\cite{WEI} to check the 
consistency of the model they proposed for describing the resonance 
broadening and shadowing 
effect in the nuclear photoabsorption. Their finding is that
above 
1~GeV a region should exist where impulse approximation is valid, i.e. the
Compton amplitudes for free and bound nucleons are the same. In addition, 
because of the influence of nucleon correlations, they predicted a small 
anti-shadowing effect at photon energies below 2~GeV, where no 
photoabsorption data
were available at the time. 

Given the new photonuclear data 
in the intermediate energy region~\cite{MIR,BIAPB}, the total
photoabsorption cross section is now well
known for a series of nuclei at photon energies up to
about~80~GeV.
This  allows
 an accurate dispersion relation analysis over the nucleonic 
resonances and shadowing threshold energy regions.
In this paper this  analysis is performed for carbon, lead, proton and 
neutron from $\omega=$~0.14 to $\omega=$~5~GeV. In section 2 we give the 
formalism used in the analysis, in section 3 we describe the procedure for the 
calculation of the FCAs, in section 4 and 5 we present and discuss 
the results for the FCA and the forward Compton scattering cross section, 
and in section 6 we recall the major findings.

\section{Formalism}

\subsection{The Forward Scattering Amplitudes}

The imaginary part of the FCA, $f_{\gamma,A}$, for photon scattering on a 
nucleus of mass number A is related to the nuclear total photoabsorption 
cross section, $\sigma_{\gamma,A}$, through the optical theorem
\begin{equation}
  Imf_{\gamma,A}(\omega)=\frac{\omega}{4\pi}\sigma_{\gamma,A}(\omega)\,.
\label{imfca}
\end{equation} 

Once the $\sigma_{\gamma,A}(\omega)$ is known for all energies,
the real part of the FCA can be calculated by using the Kramers-Kronig relation
and the well known Thomson limit~\cite{WEI}
\begin{equation}
        Ref_{\gamma,A}(\omega)=-\frac{1}{2\pi^2}\frac{Z^2}{A}S+
\frac{\omega^2}{2\pi^2}P\int_0^{\infty}\frac{\sigma_{\gamma,A}(\omega')d
\omega'}{\omega'^2-\omega^2}\,,
\end{equation}
where Z is the number of protons in the target, 
$S=\frac{2\pi^2e^2}{m}=60$~GeV $\mu$b is the value for 
the classical dipole sum,
$m$ is the nucleon mass, and $P$ denotes the principal value of the 
integral.

For energies above the pion photoproduction threshold, $\mu$, it is useful to
isolate in Eq. 2 the contribution below $\mu$~\cite{WEI} : 
\begin{equation}
        Ref_{\gamma,A}(\omega)=-\frac{1}{2\pi^2}\frac{Z^2}{A}S+
\frac{\omega^2}{2\pi^2}\int_0^{\mu}\frac{\sigma_{\gamma,A}(\omega')d
\omega'}{\omega'^2-\omega^2}+
\frac{\omega^2}{2\pi^2}P\int_{\mu}^{\infty}\frac{\sigma_{\gamma,A}(\omega')d
\omega'}{\omega'^2-\omega^2}\,,
\label{ref1}
\end{equation}
and, using the Taylor expansion of the second term on the right-hand side, 
\begin{equation}
    Ref_{\gamma,A}(\omega)\cong -\frac{1}{2\pi^2}\frac{Z^2}{A}S-
\Sigma_{\mu}+ \delta (\omega)+
\frac{\omega^2}{2\pi^2} P\int_{\mu}^{\infty}\frac{\sigma_{\gamma,A}(\omega'
)d\omega'}{\omega'^2-\omega^2},
\label{refca}
\end{equation}
where 
\begin{equation}
  \Sigma_{\mu}=\frac{1}{2\pi^2}\int_0^{\mu}\sigma_{\gamma,A}(\omega')
d\omega'
\end{equation}
and
\begin{equation}
  \delta(\omega)= \frac{-
1}{2\pi^2}\int_0^{\mu}\bigg(\frac{\omega'}{\omega}\bigg)^2
\sigma_{\gamma,A}(\omega')d\omega'
\end{equation}
are the first and third terms in the Taylor expansion.
The $\delta(\omega)$ term, which main contribution comes from the
quasi-deuteron region ($\omega=80-140$~MeV), is small compared to
$\Sigma_{\mu}$~\cite{BER,LEP,AHR3}. In this energy range and for all nuclei,  
$\frac{1}{A}\sigma_{\gamma,A}(\omega) < 
\frac{1}{A}\sigma_{\gamma,A}^{max}= 200$ 
$ \mu$b then one can write:
\begin{equation}
  |\delta(\omega)|<|\delta_{max}(\omega)|=|-
\frac{\sigma_{\gamma,A}^{max}}{6\pi^2\,A}\frac{\mu ^3}{\omega ^2}| = 3.38 \frac{\mu^3}{\omega^2} GeV \mu b 
\label{reldel}
\end{equation}
which is low for $\omega >> \mu$. In the following calculation we assumed $\delta=0$ in 
Equation~\ref{refca}, 
and evaluated the upper limit of the relevant error due to this approximation
by using the relation~\ref{reldel}.

\subsection{The Weise sum rule}

Using the dispersion relation in Equation~\ref{refca} for the nucleon and 
for the nucleus, Weise derived the following sum rule~\cite{WEI}:

\begin{equation}
\int_{0}^{\mu} d \omega \sigma_{\gamma A} (\omega) =
\frac{NZ}{A} S [ 1 + \zeta (A,Z)] \, ,
\label{wsr}
\end{equation}
where
\begin{equation}
\zeta (A, Z) = \frac{A}{NZ} \left\{5 
 \frac {R(\omega_0) + I (\omega_0)}{S} +
\left[ \frac{A_{eff} (\omega_0)}{A} - 1 \right] Z \right\}
\,,
\end{equation}
\begin{equation}
   R(\omega_0)=2\pi^2 [A_{eff}(\omega_0)Ref_{\gamma,N}(\omega_0)-
Ref_{\gamma,A}(\omega_0)]\,,
\label{delta}
\end{equation}
\begin{equation}
I(\omega_0) = \omega_0^2 \int_{\mu}^{\infty} d\omega \frac{A_{eff} (
\omega_0) -
A_{eff}(\omega)}{\omega_0^2 - \omega^2} \sigma_{\gamma N} (\omega)\,,
\label{wsrf}
 \end{equation}
$A_{eff}=\sigma_{\gamma A}/\sigma_{\gamma N}$
is the effective number of nucleons participating in the photon-nucleus
interaction,
$\omega_0$ is an energy at our disposal, large compared to 
the pion~threshold~$\mu$, 
the free nucleon FCA $f_{\gamma,N}(\omega)$,
and the photon-nucleon cross section $\sigma_{\gamma,N}(\omega)$, are
defined as an average over the numbers of protons and neutrons in the 
nucleus:
\begin{equation}
 f_{\gamma,N}(\omega)=\frac{Z}{A}f_{\gamma,p}(\omega)+
                      \frac{N}{A}f_{\gamma,n}(\omega)\,,
\end{equation}
%$f_{\gamma,p}$ and $f_{\gamma,n}$ being the proton and neutron FCA,
%respectively, and
%the photon-nucleon cross section, $\sigma_{\gamma,N}(\omega)$,
%is taken as an average over protons and 
%neutrons in the nucleus
\begin{equation}
 \sigma_{\gamma,N}(\omega)=\frac{Z}{A}\sigma_{\gamma,p}(\omega)+
                      \frac{N}{A}\sigma_{\gamma,n}(\omega)\,.
\end{equation}

\section{Calculation procedure}

%\begin{figure}
%\begin{center}\mbox{\epsfig{file=sigmamu.eps,height=10.0cm,width=12.0cm}}
%\end{center}
%\caption{Experimental values of $\Sigma_{\mu}$ for different targets. Solid
%circles are 
%results obtained by the sum of all opened channel reactions
%~\cite{LEP}
%, 
%while the open circles are measured with the transmission 
%method. Solid line is a fit to the data.
%~\cite{AHR2}
%. }
%\label{fig:figsigmu}
%\end{figure}

We performed the calculation of the real part of the FCA 
of Equation~\ref{refca}
by using the experimental
values of $\Sigma_{\mu}$ (specifically: 1.91 $NZ/A$ for carbon and 1.75 $NZ/A$
for 
lead~\cite{BER,LEP,AHR3}) and 
fits to the whole set of available data~\cite{MIR,BIAPB,MAC,PDG,HEP} for the
numerical integration
of the high energy integral.

As a fitting expression we used
the sum of a resonance and a background contributions, were the latter 
includes all non-resonant processes. Here, our main interest has not been 
focoused on the extraction of the resonance parameters, but only on the 
determination of the most faithfull mathematical description of the data.

The resonance contribution for the nucleon and the nuclei has been parametrized 
by a superposition of 
Breit-Wigner shapes~\cite{WAL}
\begin{equation}
 \sigma_{res}(s)=\sum_{i=1}^6 \sigma_i\,\frac{M_i^2 \Gamma_i^2}{(s-
M_i^2)^2+M_i \Gamma_i^2}\,,
\end{equation}
where
$\Gamma_i=\Gamma^0_i \big(\frac{k}{k_r}\big)^{2j_i}$;
 $s=2m \omega +m^2$;
$\sigma_i$, $M_i$ and $\Gamma^0_i$ are free parameters; $k$ and $k_r$ are the
center of 
mass photon momenta at the energy $\omega$ and at the resonance peak 
energy, respectively; and $j_i$ is the photon total angular momentum. The 
index $i$ refers to the six resonances considered in the fits, specifically: 
the 
$P_{33}$, $D_{13}$, $F_{15}$, $P_{11}$, $F_{37}$, $S_{11}$.

For the nucleon and for photon
energies
$\omega$ up to 4~GeV, we calculated the background contribution with the 
expression
\begin{equation}
 \sigma_{back}^{p(n)}(\omega)=(b_1^{p(n)}+\frac{b_2^{p(n)
}}{\sqrt{\omega}})\{ 1-exp[-2(\omega-\mu)]\}\,,
\end{equation}
where the indexes $p$ and $n$ refer to the proton and neutron,
and the constants are $b_1^p=91$~$\mu$b, $b_2^p=71.4$~$\mu$b GeV$^{1/2}$,
$b_1^n=87$~$\mu$b, $b_2^n=65$~$\mu$b GeV$^{1/2}$~\cite{ARM}.
Above 4~GeV we fitted the cross sections to the following expression, which 
is known to provide a good Regge behaviour~\cite{DON}
\begin{equation}
 \sigma_{back}^{p(n)}(\omega)=X^{p(n)}s^{0.0808} + 
                              Y^{p(n)}s^{-0.4525}\,,
\end{equation}
with $X^{p(n)}$ and $Y^{p(n)}$ free parameters.

For the nuclei, we calculated the background contribution with the expression
\begin{equation}
  \sigma_{back}^A(\omega)=\frac{Z}{A}\sigma_{back}^p(\omega)+
\frac{N}{A}\sigma_{back}^n(\omega)\,.
\end{equation}
To this, we added a coherent contribution, $\sigma_c$, through the
quasi-deuteron model:
\begin{equation}
 \sigma_{c}(\omega)=L\frac{NZ}{A}\sigma_{\gamma d \rightarrow np}(\omega),
\end{equation}
with the deuteron photodisintegration cross section, $\sigma_{\gamma d 
\rightarrow np}(\omega)$, calculated by using the fit of 
Rossi et al.~\cite{ROS}, and the Levinger constant values $L$ given by
Tavares et al.~\cite{TAV}. In addition,
to include the shadowing effect, we 
multiplied the cross section for nuclei by the factor  
\begin{equation}
  h(\omega)=1-K_1 exp(-K_2/\omega^2)-K_3 exp(-K_4/\omega^2)\,,
\end{equation}
with the constants $K_i$ (i = 1,4), different for each nucleus, being
 free parameters in the fit.

\begin{figure}
\begin{center}\mbox{\epsfig{file=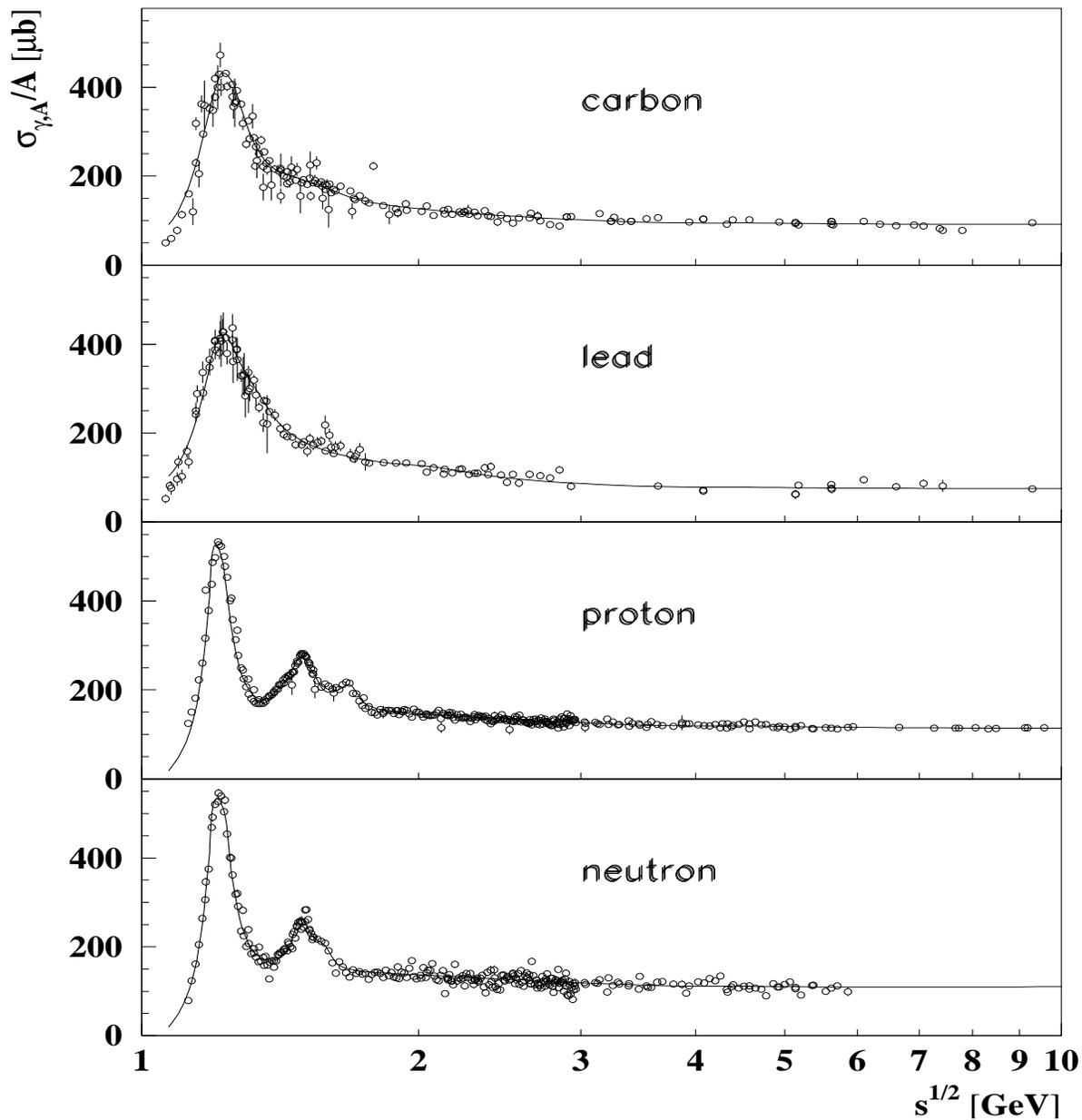,height=17.0cm,width=17.0cm}}
\end{center}
\caption{Data for the total photoabsorption cross section and the relevant 
fits used in the calculation of the FCA for the given nuclei
.}
\label{fig:figfits}
\end{figure}

In Figure~\ref{fig:figfits} we compare the fits with the
data: the reduced $\chi^2$ are 0.37, 0.13, 0.4 and 0.22 for C, Pb, p and n, 
respectively. The data for proton are from 
reference~\cite{PDG}, those for carbon and lead are the total photo-hadronic 
cross sections from the HEPDATA 
compilation~\cite{HEP} and from
references~\cite{MIR,BIAPB}. The data for the neutron were obtained 
from the deuteron ones~\cite{PDG} by using the procedure described in 
references~\cite{BIAPB,BAB}.

\section {The forward Compton amplitudes}

\begin{figure}
\begin{center}\mbox{\epsfig{file=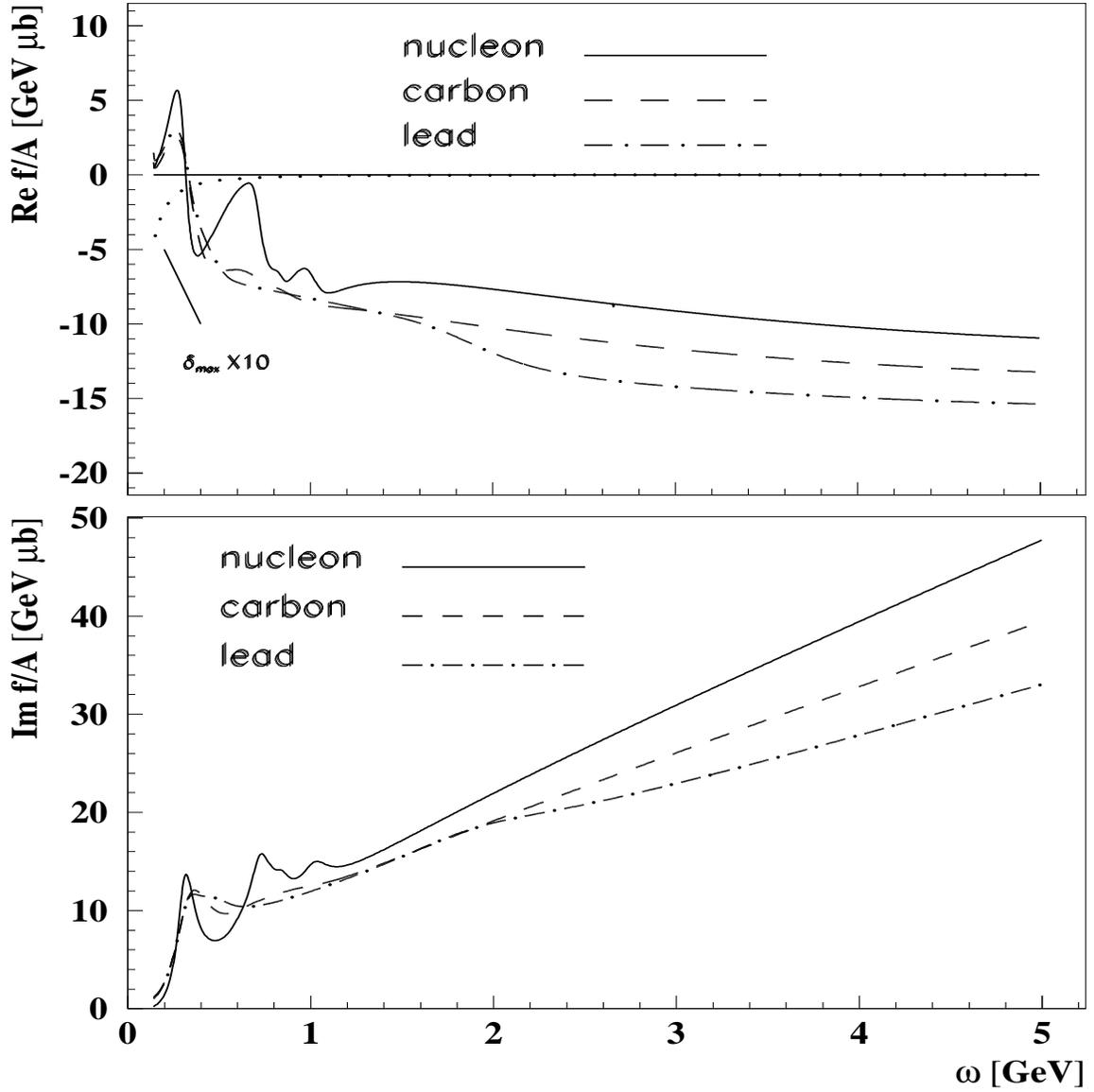,height=17.0cm,width=17.0cm}}
\end{center}
\caption{The FCA for the given nuclei as a function of the photon
energy. The curve for the free-nucleon is calculated from Eq. 2
assuming Z=N=A/2. The dotted line represents the $\delta_{max}$ contribution 
(multiplied by a factor 10) 
of Eq.~(7).
 }
\label{fig:figfca}
\end{figure}

The real and imaginary parts of the FCA were calculated from 
Equations~(\ref{imfca}) and~(\ref{refca}) by numerical integration of the
fits to the photoabsorption cross sections. The results are 
shown in Fig.~\ref{fig:figfca}.
In order to give an estimate of
the error due to the approximation $\delta $~=~0 used in Eq.~\ref{refca}, 
the quantity 
$\delta_{max}(\omega)$ is also shown in the Figure. As it is seen, 
$|\delta_{max}(\omega)|<0.1$~$ GeV \mu$b
for $\omega>0.3$~GeV
and therefore the used approximation is reliable above that energy.
For the sake of comparison with the free-nucleon, the nuclear FCAs have 
been divided by the relevant mass numbers. Henceforth the shortned form FCA
expresses this quantity.

%\begin{figure}
%\begin{center}\mbox{\epsfig{file=ref.eps,height=10.0cm,width=12.0cm}}
%\end{center}
%\caption{The real part of the FCA normalized to the mass number A as a 
%function of the photon energy for 
%different targets, together with the quantity $\delta_{max}(\omega)$ 
%that was evaluated by relation (7). The curve for the 
%free-nucleon was calculated assuming Z=N.}
%\label{fig:figreal}
%\end{figure}

%\begin{figure}
%\begin{center}\mbox{\epsfig{file=imf.eps,height=10.0cm,width=12.0cm}}
%\end{center}
%\caption{The imaginary part of the FCA normalized to the mass number A as a 
%function of the photon energy
%for different targets. The curve for the free-nucleon is calculated 
%assuming Z=N.}
%\label{fig:figima}
%\end{figure}

%The uncertainties in the FCA coming from the fits and from the integration 
%process increase approximately linearly from 0 at 0.14~GeV to 9\% at 
%1~GeV, and remain approximately constant between 1 and 5~GeV.

Both the real and the imaginary FCA curves for 
the free-nucleon and for the nuclei have similar shapes in the 
$\Delta$-resonance region, reflecting clearly the $\Delta$-resonance
excitation in the nucleus. The different $\Delta$-resonance
width, which is more clearly seen in the $Imf(\omega)$ curves, is 
due to Fermi motion and to $\Delta$-resonance propagation in nuclear matter.

On the contrary, above the $\Delta$-resonance region and up to
$\omega \simeq 1.2$~GeV, they are quite different:
while the signatures of the high-mass nucleon 
resonances are evident in the curves for the free-nucleon, 
they are absent in those for nuclei. This difference can not be attributed 
to the smearing due to the 
integration over the Fermi momentum only~\cite{VAL}.
 Moreover, the absolute values of the real and imaginary FCA curves 
for the bound-nucleon are significantly lower than
those for the free-nucleon. The latter effect persists at higher energies where
the resonance contribution vanishes. This fact may 
reflect the onset of the shadowing, as suggested in ref.~\cite{MIR}.

%The small uncertainties in the FCA in the $\Delta$ region allow the 
%comparison between the curves for carbon and lead.The different resonance
%width, which is more clearly seen in the $Imf(\omega)$ curves, is 
%due to Fermi motion and to resonance propagation in nuclear matter.

The energies $\omega^*$ at which $Ref(\omega^*)=0$ (henceforth called 
zeros of $Ref$), that correspond to the
maximum in the photoabsorption cross sections,
give an indication  of the $\Delta$-resonance peak 
positions.
For the proton one finds  $\omega^*_p = 0.317 \pm 0.002$ GeV. 
In the Breit-Wigner parametrization of the  $\Delta$-resonance given 
by Walker et al.~\cite{WAL} this result is  in good agreement with
 the $\Delta$-mass value
$M_\Delta = 1.232$ GeV. This finding suggests that, in spite of the
presence of a non-resonant background the zero of $Ref$ is a meaningful
indication of the $\Delta$-peak position.
For carbon and lead one finds  $\omega^*_C = 0.328 \pm 0.002$ GeV and 
$\omega^*_{Pb} = 0.332 \pm 0.002$ GeV. Assuming that the non-resonant
contribution is not very different for nuclei as compared to the 
nucleon, this
 confirms the
increase  of the $\Delta$ mass inside 
nuclei suggested in previous analyses~\cite{BIAPB,BOF2}.

%\begin{table}[th]
%\caption{The energies $\omega^*$ where $Ref_{\gamma,A}(\omega^*)=0$ for proton%, carbon and lead}
%\label{tb-refz}
%\newcommand{\GeV}{\mathrm{GeV}}
%\newcommand{\Rule}{\rule[-.7ex]{0ex}{2.9ex}}
%\begin{center}
%\begin{tabular}{lll} \hline
%\Rule Nucleus & $\omega^*$ [GeV]    \\ \hline
%\Rule         & $(\GeV)$        \\ \hline
%\Rule  p       & $0.317\pm0.002$ \\ 
%\Rule  C      & $0.328\pm0.002$ \\ 
%\Rule  Pb     & $0.332\pm0.002$ \\ \hline
%\end{tabular}
%\end{center}
%\end{table}

\begin{figure}
\begin{center}\mbox{\epsfig{file=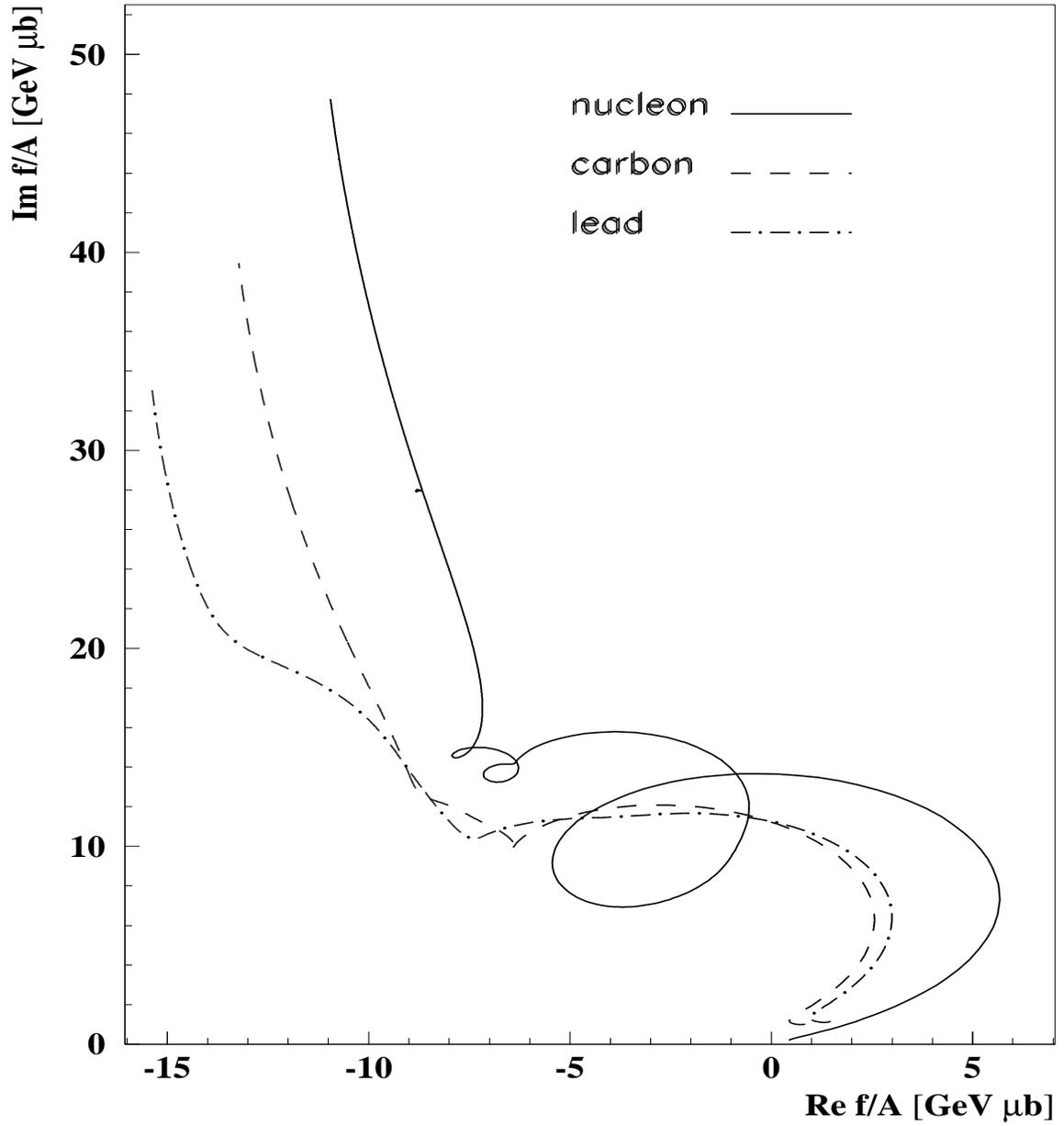,height=17.0cm,width=17.0cm}}\end{center}
\caption{The Argand plot of the FCAs for the given nuclei. 
The curve for the free-nucleon was calculated assuming Z=N.}
\label{fig:figarg}
\end{figure}

The differences in the FCA for free and bound nucleons
 are more cleanly seen in the Argand plots shown
in Figure~\ref{fig:figarg}. While in the free-nucleon curve one sees the 
loops that reflect the presence of the resonance peaks in the cross section,
for carbon and lead there is only the loop corresponding to the 
$\Delta$-resonance, while 
those for the high energy resonances disappear.
The curve for carbon shows a cusp just after the 
$\Delta$-loop, at $\omega=0.525$~GeV corresponding to the minimum in 
the imaginary part of FCA.
Also for lead the main deviation
from the free-nucleon behavior starts around 0.525~GeV.
This deviation can be ascribed to the different relative strength of the
resonance and 
background contributions in nuclei respect to the nucleon case,
as recently suggested~\cite{GIA,OSE}.

\section{The forward Compton scattering cross section}

\begin{figure}
\begin{center}\mbox{\epsfig{file=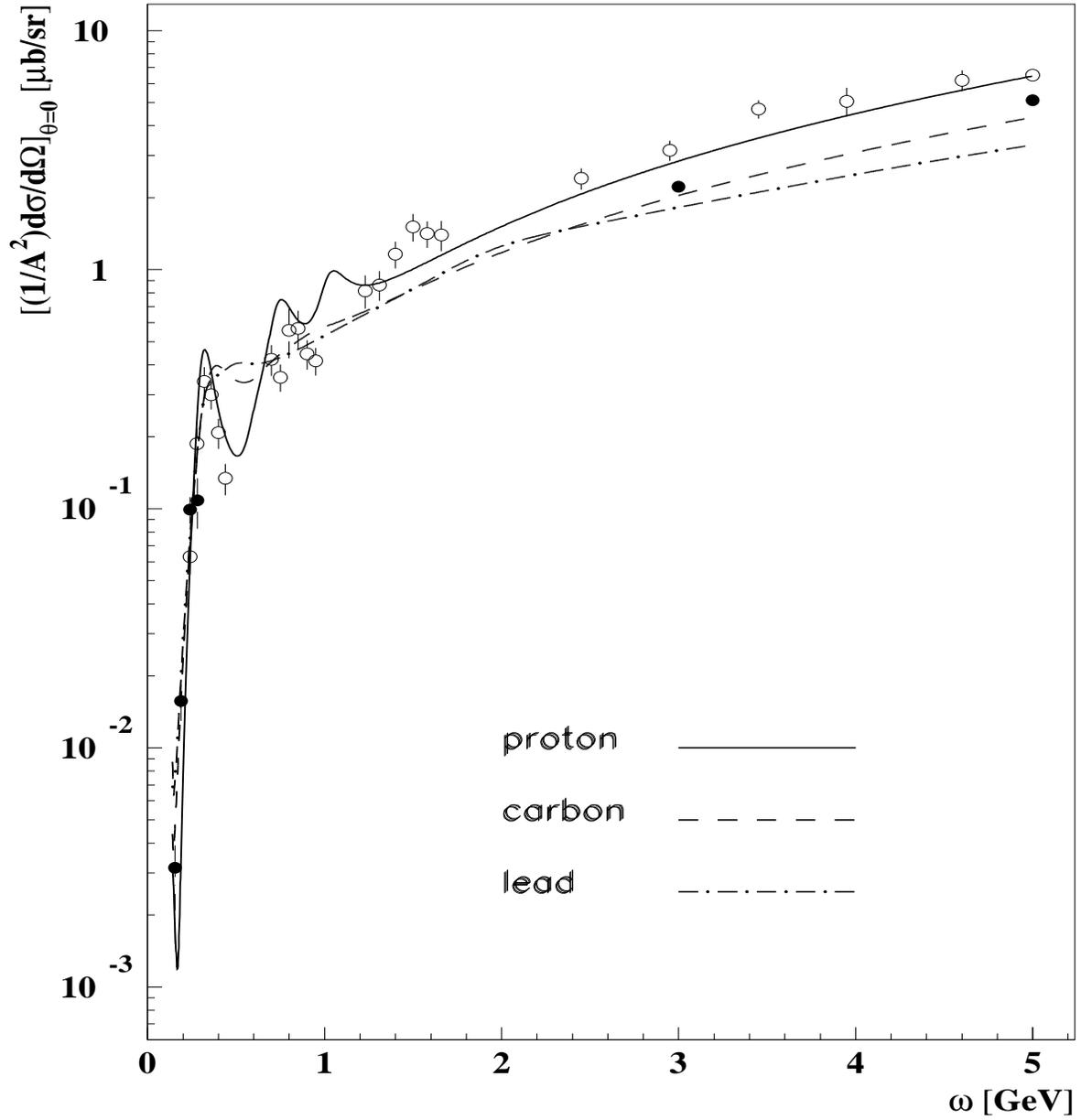,height=17.0cm,width=17.0cm}}
\end{center}
\caption{Forward Compton scattering differential cross section as 
calculated with the FCA worked out is this paper. Open and closed circles 
are data for proton [24, 25, 26, 27] and carbon [28, 29].}
\label{fig:figelascat}
\end{figure}

Using the above determined FCAs, we have calculated the forward Compton
scattering cross sections for proton,
carbon and lead 
\begin{equation}
   \frac{d\sigma_{\gamma A \rightarrow \gamma A}}{d\Omega}(\theta=0,\omega)
=
                    |f_{\gamma,A}(\omega)|^2\,.
  \label{elascat}
\end{equation}
The 
results are shown in Figure~\ref{fig:figelascat}:
as can be seen, the
difference between bound and free nucleons is 
enhanced since the nuclear Compton cross section is equal to 
the sum of the squares of both the real and imaginary parts.
The high-mass nucleon-resonances are clearly seen in the proton
curve, but are absent in the nuclear ones.

Also shown in the Figure are the 
available data for proton~\cite{GEN,JUN, DUD, CRIEP} and  
carbon~\cite{IGA,CRIEC}.
Unfortunately, there are no data for lead and the interesting 
region of the  nucleon-resonances 
is uncovered by the experiments for carbon.
 It is worth noticing that 
proton data at energies between 0.6 and 1.6  GeV
 are obtained by extrapolating 
from Compton scattering measurements at relatively high scattering
angles ($ \Theta_{c.m.} \geq$ 35 deg)~\cite{GEN,JUN, DUD} and then 
suffer for larger (up to $\pm 22$ \%) 
systematic uncertainties (not shown in Fig.~\ref{fig:figelascat}), due
to the extrapolation procedure. 
The overall good agreement found between the curves for 
proton and carbon and the corresponding data demonstrates the 
reliability of
the FCA values calculated with Equations~\ref{imfca} and~\ref{refca}.
In addition, this agreement shows the consistency between two complementary
sets of data: specifically, 
the total photoabsorption and the forward Compton scattering data.

This validation of the FCAs is also a verification of
the Weise sum rule, which is trivially satisfied by the 
FCAs
calculated 
here, as can be easily verified by direct substitution of 
Equation (~\ref{refca}) into Equation~(\ref{wsr}).
%Such a verification has already been performed for the proton by 
%Alvensleben~\cite{ALV}, but 
%has never been done for the nuclei.

The FCA curves are also a useful tool for testing the validity of the
models recently proposed to describe the damping of high-mass 
nucleon-resonances in nuclear medium~\cite{GIA,EFF}, and the shadowing effect
at high energy~\cite{BOFFI}.

\begin{figure}
\begin{center}\mbox{\epsfig{file=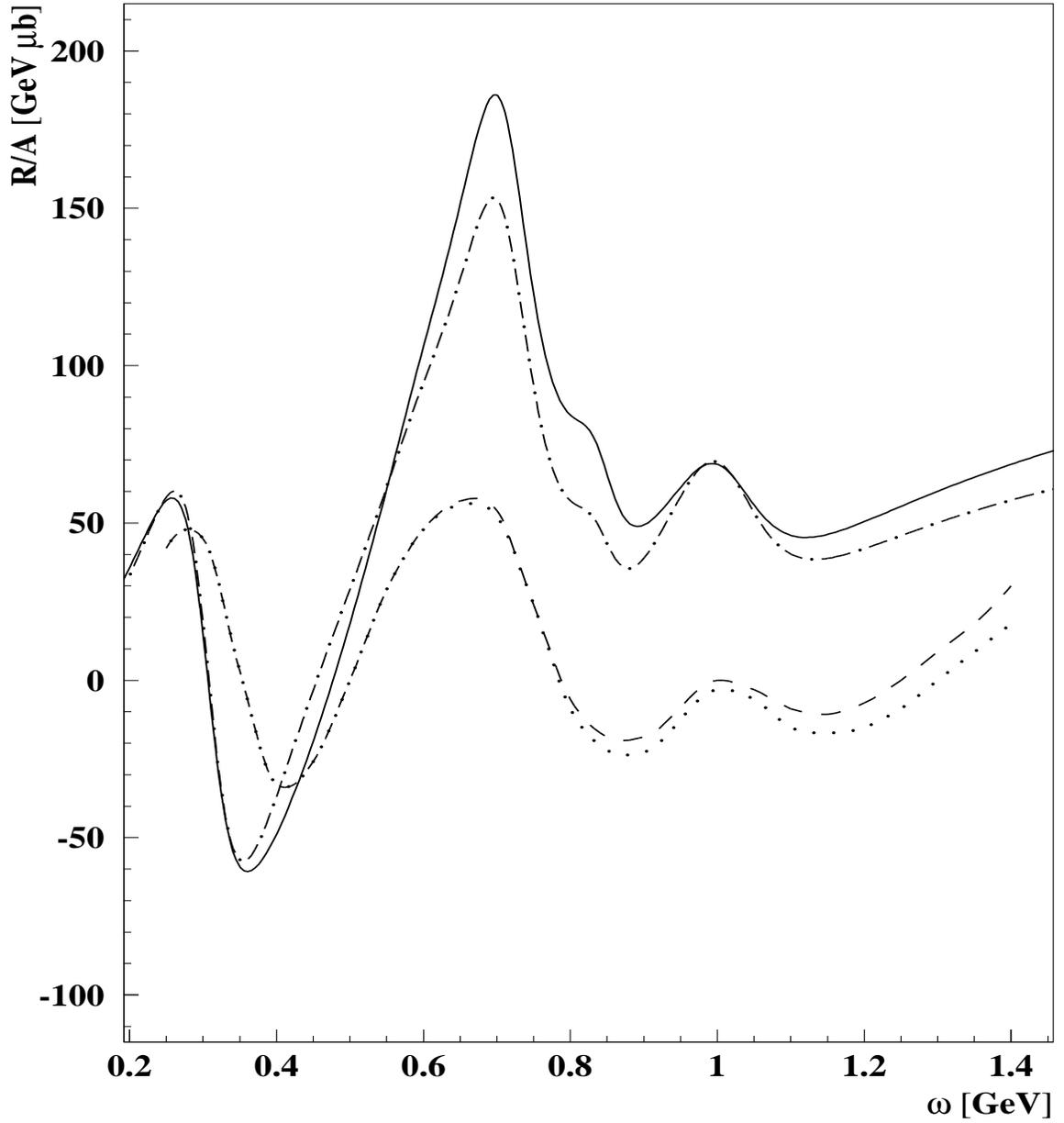,height=17.0cm,width=17.0cm}}
\end{center}
\caption{The quantity $R(\omega)$ defined in Equation 10 calculated 
for lead (solid line) and carbon (dashed-dotted line) is compared with 
the results for uranium both in the case of correlation (dotted line)
and without correlation (dashed line) from reference [5].}
\label{fig:figbof}
\end{figure}

As an application, we examine the analysis performed by Boffi et 
al.~\cite{BOFFI} who evaluated the quantity $R(\omega)$ of 
Equation~(\ref{delta}) for uranium. In Figure~\ref{fig:figbof} we show 
their predictions
together with our results for carbon and  lead.
As it is seen the two calculations largely disagree
in the whole energy range, and the disagreement is stronger than 
expected from mass number dependence.
In the $\Delta$-region it arises from the less accurate Breit-Wigner 
parametrization of the nucleon-resonances used by Boffi et al., which 
results in a shift of the 
resonance-mass. At higher energies it is due to
a not accurate accounting of the shadowing 
contribution which produces an overestimate of the new photoabsorption data.
The disagreement is still larger with the calculation where 
 the two-nucleon correlations are included,  which 
give  an  
antishadowing effect for $\omega$~$<$~2~GeV.

%The clear disagreement between the two curves indicates that the
%calculation of Boffi et al.
%is not correct, and invalidates the model they proposed to describe the 
%damping of the high-mass nucleon-resonances as a result of the interference 
%between the resonant and non resonant contributions to the scattering 
%amplitudes, and the shadowing effect at high energy as a result of the 
%coherent sum of the individual nucleon amplitudes.

\section{Conclusion}

We have used the data set of the total photoabsorption cross 
section  on carbon and lead, now available in the full range from 0.14 to 80
GeV,  to perform an accurate dispersion 
relation analysis and to determine the forward Compton scattering amplitudes 
and cross section. 
The results of this analysis show that
\begin{itemize}

\item[i)] in the $\Delta$-resonance region, the FCA curves for the nuclei 
are similar in shape to the free nucleon ones and the
locations of the zero of $Ref_{\gamma A}(\omega)$ suggest a  shift
of the $\Delta$-resonance mass inside nuclei, in agreement with the 
conclusions
previously drawn from a different analysis~\cite{BOF2};

\item[ii)] in the high-mass nucleon resonance region, the FCA curves
 for the nuclei
strongly differ
from the free-nucleon one. The FCA Argand plots for the nuclei 
show a sharp deviation from the corresponding 
curve for the free-nucleon  in between of 0.5~-~0.6~GeV, with no signature of
high-mass nucleon resonances;

\item[iii)] above the nucleon resonance region the reduction of the FCAs for the
nuclei may reflect the onset of the shadowing effect;

\item[iv)]  the forward Compton scattering cross sections derived from the 
FCAs for carbon and 
proton are in  good agreement with data.

\end{itemize}

\pagebreak


\begin{thebibliography}{MIR}
\bibitem{AHR} J. Ahrens, L. S. Ferreira and W. Weise, Nuc. Phys. A485 (1988
) 621.
\bibitem{BIAPB} N. Bianchi et al., Phys. Rev. C54 (1996) 1688
\bibitem{HAT}T. Hatsuda and T. Kunihiro, Phys. Report 247 (1994) 221, and reference therein.
\bibitem{MUK} N. C. Mukhopadhyay and V. Vento, RPI Internal Report RPI-97,1997-
N117.
\bibitem{BOFFI} S. Boffi, Ye. Golubeva, L. A. Kondratyuk and M. I.
Krivoruchenko, Nuclear Physics A606 (1996) 421.
\bibitem{WEI} W. Weise, Phys. Rev. Lett. (1973) 773 and W. Weise, Phys. Rep
. 13 (1974) 53.
%\bibitem{AHR2} J. Ahrens, Nucl. Phys. A446 (1985) 229c, and reference
%therein.
\bibitem{MIR} M. Mirazita et al., Phys. Lett. B407 (1997) 225.
%\bibitem{BIAC} N. Bianchi et al., Phys. Lett. B309 (1993) 5.
%\bibitem{BIAPB} N. Bianchi et al., Phys. Rev. C54 (1996) 1688.
\bibitem{BER} B. L. Berman, R. Berg\`ere and P. Carlos, Phys. Rev. C26 (
1982) 304.
\bibitem{LEP} A. Lepr\`etre et al., Nuc. Phys. A367 (1981) 237.
\bibitem{AHR3} J. Ahrens, Nucl. Phys. A251 (1975) 479.
\bibitem{MAC} M. Mac Cormick et al, Phys. Rev. C53, (1996) 41.
\bibitem{PDG} Particle Data Group (R.M. Barnett et al.), Phys. Rev. D54, (
1996) 191.
\bibitem{HEP} The Durham RAL Databases, http://DUR\-PDG.DUR.AC.UK/HEP\-DATA.
\bibitem{WAL} R. L. Walker et al., Phys. Rev. 182 (1969) 1729.
\bibitem{ARM} T. A. Armstrong et al., Phys. Rev. D5 (1972) 1640.
\bibitem{DON} A. Donnachie and P.V. Landshoff, Phys. Lett. B296 (1992) 227.
\bibitem{ROS} P. Rossi et al., Phys. Rev. C40 (1989) 2412.
\bibitem{TAV} O. A. P. Tavares and M. L. Terranova, J. Phys. G18 (1992) 521.
%\bibitem{HEP} http://DURPDG.DUR.AC.UK/HEPDATA.
\bibitem{BAB} D. Babusci, G. Giordano and G. Matone, Phys. Rev. C57 (1998) 
291.
\bibitem{VAL} V. Muccifora Nucl. Phys. A623 (1997) 116c.
%\bibitem{VMD} T. H. Bauer, R. D. Spital and D. R. Yennie, Rev. Mod. Phys.
%50 (1978) 261.
\bibitem{BOF2}  S. Boffi, Ye. Golubeva, L. A. Kondratyuk, M. I.
Krivoruchenko and E. Perazzi, Phys. of Atomic Nuclei, 60 7 (1997) 1193.
\bibitem{GIA} M.M Giannini and E.Santopinto, Phys. Rev. C  49,3 (1994) 
R1258.
\bibitem{OSE} J. A. G\'omez Tejedor, M. J. Vicente-Vacas and E. Oset, Nucl.
Phys. A588 (1995) 819.
\bibitem{GEN} H. Genzel et al., Z. Phys. A279 (1976) 399.
\bibitem{JUN} M. Jung et al., Z. Phys. C10 (1981) 197.
\bibitem{DUD} J. Duda et al., Z. Phys. C17 (1983) 319.
\bibitem{CRIEP} L. Criegee et al., Nuc. Phys. B121 (1977) 31.
\bibitem{IGA} R. Igarashi et al., Phys. Rev. C52 (1995) 755.
\bibitem{CRIEC} L. Criegee et al., Nuc. Phys. B121 (1977) 38.
\bibitem{EFF} M. Effenberger et al., Nucl. Phys. A613 (1997) 353.

%\bibitem{ALV} H. Alvensleben et al., Phys. Rev. Lett. 30 (1973) 328.
%\bibitem{BIAU} N. Bianchi et al., Phys. Lett. B299 (1993) 219.
%\bibitem{GGT} M. Gell-Mann, M. Goldberger and W. Thirring, Phys. Rev. 95 (
%1954)
%\bibitem{DALI} R. H. Dalitz in "Strange Particles and Strong Interactions",
%Oxford Univ. Press, 1962, chapter XI.

\end{thebibliography}
\end{document}